\documentclass[conference]{IEEEtran}
\IEEEoverridecommandlockouts

\usepackage{multirow}
\usepackage{hhline}
\usepackage{verbatim}
\usepackage{cite}
\usepackage{enumitem}
\usepackage{algorithm}
\usepackage{algpseudocode}
\usepackage{graphics} 
\usepackage{graphicx}
\usepackage{subfig}
\usepackage{url}
\usepackage{balance}
\usepackage{siunitx}
\usepackage{booktabs}
\usepackage[font={small}]{caption}
\usepackage{amsmath,amssymb,amsfonts}

\newcommand{\R}{\mathbb{R}}

\usepackage{hyperref} 
\begin{document}

\title{Contrastive Representation Learning for Predicting Solar Flares from Extremely Imbalanced Multivariate Time Series Data}

\author{\IEEEauthorblockN{Anonymous}}

\author{
    \IEEEauthorblockN{Onur Vural, Shah Muhammad Hamdi, Soukaina Filali Boubrahimi}
    \IEEEauthorblockA{Department of Computer Science, Utah State University, Logan, UT, USA\\
    Email: a02429549@aggies.usu.edu, hamdi@usu.edu, boubrahimi@usu.edu}
}

\author{
    \IEEEauthorblockN{Onur Vural, Shah Muhammad Hamdi, Soukaina Filali Boubrahimi}
    \IEEEauthorblockA{Department of Computer Science, Utah State University, Logan, UT 84322, USA\\
    Emails: \{onur.vural, s.hamdi, soukaina.boubrahimi\}@usu.edu \\
    \href{https://orcid.org/0009-0004-4950-7520}{ORCID: 0009-0004-4950-7520}, \href{https://orcid.org/0000-0002-9303-7835}{0000-0002-9303-7835}, \href{https://orcid.org/0000-0001-5693-6383}{0000-0001-5693-6383}}
}

\maketitle

\begin{abstract}
Major solar flares are abrupt surges in the Sun's magnetic flux, presenting significant risks to technological infrastructure. In view of this, effectively predicting major flares from solar active region magnetic field data through machine learning methods becomes highly important in space weather research. Magnetic field data can be represented in multivariate time series modality where the data displays an extreme class imbalance due to the rarity of major flare events. In time series classification-based flare prediction, the use of contrastive representation learning methods has been relatively limited. In this paper, we introduce CONTREX, a novel contrastive representation learning approach for multivariate time series data, addressing challenges of temporal dependencies and extreme class imbalance. Our method involves extracting dynamic features from the multivariate time series instances, deriving two extremes from positive and negative class feature vectors that provide maximum separation capability, and training a sequence representation embedding module with the original multivariate time series data guided by our novel contrastive reconstruction loss to generate embeddings aligned with the extreme points. These embeddings capture essential time series characteristics and enhance discriminative power. Our approach shows promising solar flare prediction results on the Space Weather Analytics for Solar Flares (SWAN-SF) multivariate time series benchmark dataset against baseline methods.
\end{abstract}


\section{Introduction} \label{sec:intro}
In the domain of solar activity, a solar flare appears as a sudden surge in magnetic flux, originating from the Sun's surface and extending into the solar corona and heliosphere. Classified logarithmically based on the peak soft X-ray flux in the 1–8 Å wavelength range, flares are denoted by categories A, B, C, M, and X where M and X classes suggest intense flare activity \cite{ahmadzadeh2021train}. Such solar phenomena may emit gamma-ray, x-ray, and extreme ultraviolet radiation, posing radiation-induced risks to astronauts, technological infrastructure, electronic devices, navigation, and communication systems \cite{eastwood2017economic}. The 1859 Carrington Event demonstrates the potential magnitude of a solar superstorm's impact, with a recurrence potentially causing prolonged blackouts and massive economic losses in our technology-dependent society \cite{angryk2020multivariate}. Consequently, the heliophysics community underscores the importance of meticulous data analysis and the exploration of diverse methods for robust predictions of major flares from solar active region magnetic fields and flare data.

In 2020, the Space Weather Analytics for Solar Flares (SWAN-SF) dataset \cite{angryk2020multivariate} was introduced as a pivotal resource in solar flare research. Derived from Space-weather HMI Active Region Patch (SHARP) solar photospheric vector magnetograms, SWAN-SF includes multivariate time series (MVTS) data from May 2010 to December 2018. It features 24 flare-predictive magnetic field parameters and over 10,000 flare reports, reformulating solar flare prediction as an MVTS classification task. Recent models built upon SWAN-SF MVTS data have shown enhanced efficacy in classifying flaring activities compared to earlier single timestamp models \cite{angryk2020multivariate}. Due to major flares occurring rarely, the extreme class imbalance is a remarkable challenge in SWAN-SF. Accordingly, an effective learning methodology is representation learning, the process of learning meaningful fixed-dimension embeddings as data representations from the raw input data domain that keep their inherent features and have better transferability in downstream tasks. However, in the time series domain and particularly in solar flare tasks, the use of contrastive representation learning methods has been relatively limited. In contrastive methods, positive samples are contrasted with negative samples such that similar examples are mapped closer in the new feature space while the distance between dissimilar examples is maximized. Contrastive learning methods have been widely adopted in various domains for their soaring performance in representation learning, including vision, language, and graph-structured data \cite{le2020contrastive}. Consequently, it becomes imperative to explore the impact of using contrastive approaches with the ultimate goal of enhancing flare prediction performance.

In this paper, we introduce CONTREX, a novel contrastive representation learning approach designed specifically for extremely imbalanced time series characteristics of data points. Our method consists of four main parts: first, we extract features from MVTS instances that capture their essential dynamical properties. Next, we derive two contrastive representations from the feature vectors of positive and negative instances that provide maximum separation capability. Then, we train a sequence representation embedding module with the original MVTS instances to generate embeddings that encapsulate time series characteristics, guided by our custom contrastive reconstruction loss. Finally, based on these embeddings, we employ a downstream classifier for binary classification. This comprehensive method enables effective representation learning and classification in time series analysis tasks. The contributions made by this paper are listed as: 

\begin{itemize}[leftmargin=*]
    \item Introducing a novel contrastive learning framework tailored specifically for time series data that can be applied to both univariate and multivariate settings to address the challenges posed by temporal dependencies and extreme class imbalance.
    \item Defining a custom contrastive reconstruction loss during training of the sequence representation embedding module to guide the embeddings towards the positive and negative extremes, enhancing the discriminative power of the learned representations.
    \item Experimentally demonstrating the effectiveness of the proposed method regarding a performance metric evaluation reflecting the nature of the benchmark dataset.   
\end{itemize}

\section{Related Works} \label{sec:Related Work}

\subsection{Solar Flare Prediction}
One of the earliest endeavors in solar flare prediction history was THEO, an expert system that relied on human entries, officially used by The Space Environment Center (SEC) of the National Oceanic and Atmospheric Administration (NOAA) in 1987 \cite{mcintosh1990classification}. Subsequently, as space-based and ground-based observatories amassed a wealth of magnetic field data, flare prediction transitioned into a data science task with models emerging based on line-of-sight and vector magnetograms, delineating solar active region photospheric magnetic field parameters. Since 2010, NASA's Solar Dynamics Observatory (SDO) has been continuously mapping the full-disk vector magnetic field every 12 minutes through the Helioseismic and Magnetic Imager (HMI) instrument, leading to reliance on this continuous vector magnetogram data in current literature \cite{mason2010testing}. Nonlinear statistical models, particularly machine learning classifiers have gained prominence in solar flare prediction such as logistic regression \cite{song2009statistical}, C4.5 decision tree \cite{yu2009short}, fully connected neural network \cite{ahmed2013solar}, support vector machine \cite{bobra2015solar}, and relevance vector machine \cite{al2015automated}. The introduction of temporal window-based flare prediction \cite{angryk2020multivariate} led to the creation of the SWAN-SF dataset \cite{DVN/EBCFKM_2020}, which records magnetic field data over time. Following SWAN-SF, various MVTS classification methods emerged, including kNN training with statistical summarization \cite{hamdi2017time}, MVTS decision trees with clustering \cite{ma2017solar}, LSTM-based sequence modeling \cite{muzaheed2021sequence}, and functional network embedding \cite{hamdi2022multivariate}. These advancements signify a shift from traditional linear models to sophisticated machine learning techniques in predicting solar flare activities.

\subsection{Time Series Contrastive Representation Learning}
In contrastive representation learning, unlike learning a mapping to labels as in discriminative models or reconstructing input samples as in generative models, data representations are learned through comparison between data points in the input space by mapping similar data points close while increasing the distance between dissimilar data points in the embedding space \cite{le2020contrastive}. Although contrastive learning methods are less explored in the time series domain than vision, language, and graph domains, there is growing interest. Accordingly, sampling positive and negative instances and time pieces from the anchor to learn inter-sample and intra-temporal relations \cite{fan2020self}, temporal and contextual contrasting by creating two views for each sample with strong and weak augmentations \cite{eldele2021time}, introducing fidelity and variety criteria, and creating a meta-learner for selecting feasible data augmentations \cite{luo2023time}, improving representation quality by instance-wise and temporal contrastive loss with soft assignments \cite{lee2023soft}, employing a siamese structure and convolutional encoder to learn representations without negative pairs \cite{zheng2023simts} were such works that highlight the potential of contrastive learning in enhancing time series representation learning.

\section{Methodology} \label{sec:Methodology}

\subsection{Extracting Dynamical Time Series Features}

To capture the dynamical properties of time series, catch22 feature extraction method \cite{lubba2019catch22} is selected. The 22 features obtained by catch22 method give a low dimensional summary to represent the diverse and interpretable characteristics of time series including linear and non-linear autocorrelation, successive differences, value distributions and outliers, and fluctuation scaling properties. The catch22 method has shown highly discriminative and low redundancy feature representation power for various benchmark time series datasets \cite{ruiz2021great}. In our approach, we extract catch22 features for each univariate time series in MVTS instances. Accordingly, for each MVTS data instance $M^{\text{(k)}} \in \mathbb{R}^{\tau \times N}$ having $N$ parameters as univariate time series with a length of $\tau$, we extract a fixed-dimensional multi-catch22 vector $V^{\text{(k)}} \in \mathbb{R}^{D}$ where $D$, the length of fixed-dimensional multi-catch22 vector, is equal to $22N$.

\subsection{Obtaining Contrastive Extremes}
After obtaining $K$ fixed-dimensional multi-catch22 vectors $V^{\text{(k)}} \in \mathbb{R}^{D}$, $K$ being total number of MVTS instances, the second step in our proposed method is to obtain two extreme points as overarching representations for positive and negative classes to enhance the contrastive power, effectively drawing positive data points closer to the positive extreme and negative data points closer to the negative extreme. Positive and negative extremes $E_{P} \in \mathbb{R}^{D}$ and $E_{N} \in \mathbb{R}^{D}$ are selected as multi-catch22 vectors that yield the complete linkage, representing the data points that yield the greatest distance between clusters:

\begin{equation}
D(P, N) = \max_{V_{P} \in P, V_{N} \in N} d(V_{P}, V_{N}) \label{eq:extremes}
\end{equation}

In (\ref{eq:extremes}), \( D(P, N) \) is the distance between clusters of positive and negative classes, and \( d(V_{P}, V_{N}) \) is the Euclidian distance between positive multi-catch22 vector $V_{P} \in \mathbb{R}^{D}$ and negative multi-catch22 vector $V_{N} \in \mathbb{R}^{D}$.  

\subsection{Framework}
\begin{figure*}[t]
\centering
\includegraphics[width=0.64\linewidth]{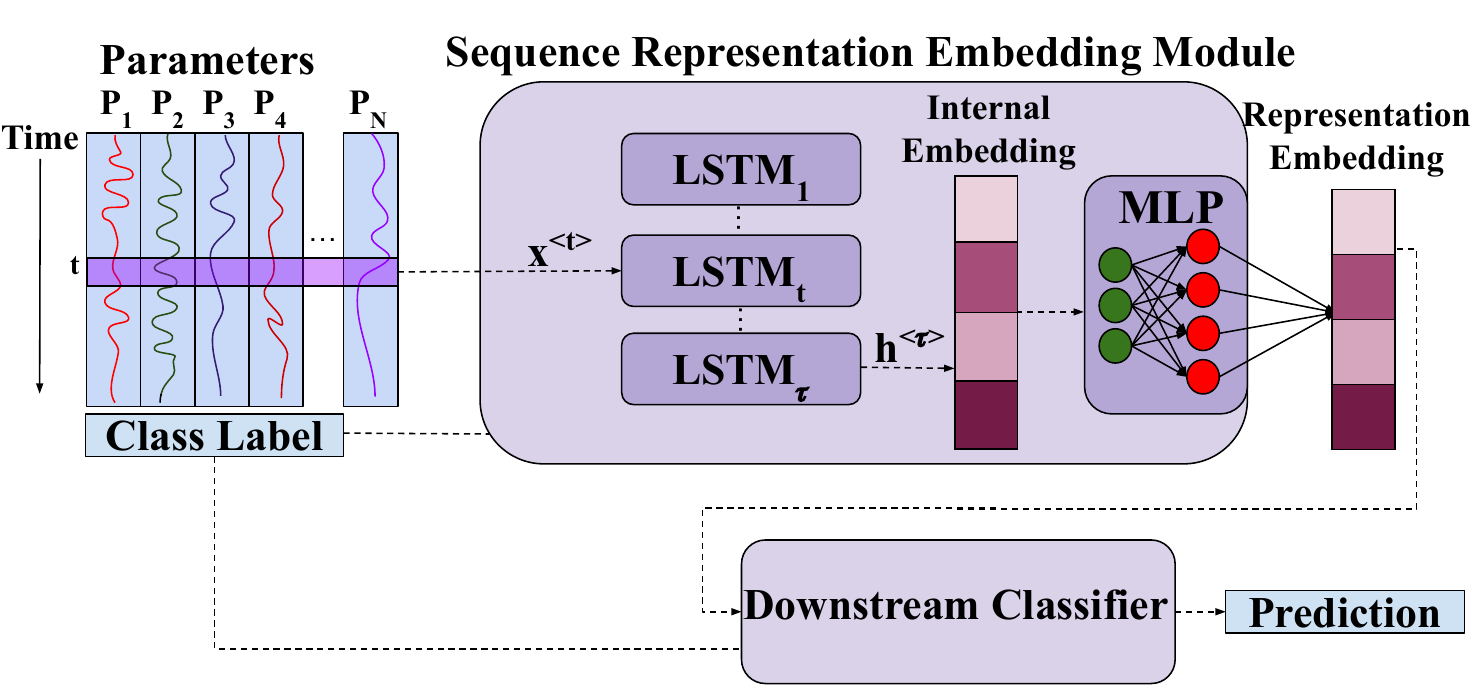}
\caption{CONTREX is composed of (1) a sequence representation embedding module to derive fixed-dimensional embeddings from MVTS data points, and (2) a downstream classifier that utilizes the representation embeddings for binary prediction.}
\label{fig:framework}
\end{figure*}

CONTREX is composed of two modules integrated in an end-to-end manner. The initial module derives fixed-dimensional embeddings from MVTS data points, while the subsequent module leverages these embeddings to execute the binary classification task. Fig. \ref{fig:framework} provides an overview of our architecture.

\subsubsection{Obtaining Representation Embeddings}
In the sequence representation embedding module, the sequence modeling of MVTS data points is done via a long short-term memory (LSTM) layer where each MVTS instance $M^{\text{(k)}} \in \mathbb{R}^{\tau \times N}$ is regarded as total $\tau$ timestamp vectors $x^{<t>} \in \R^{N}$, which are sequentially processed by the LSTM cells. The input size corresponds to $N$ parameters, and the final hidden state representation $h^{<\tau>}$ outputs an internal embedding vector. Subsequently, this vector undergoes projection to a D-dimensional vector, mirroring the size of multi-catch22 vectors, facilitated by a multilayer perceptron (MLP) layer. Our network incorporates a single dropout layer to enhance robustness. After the training is complete, for the $k^{\text{th}}$ MVTS instance, the embedding vector $X_{label}^{\text{(k)}} \in \mathbb{R}^{D}$ can be extracted from the last layer.

\subsubsection{Loss Function}
Here, we propose a novel loss function, the contrastive reconstruction loss to guide the training of our sequence representation embedding module such that it will learn similar representations to the extremes in a supervised setting. Accordingly, for a positive class MVTS instance $X_{P}^{\text{(k)}} \in \mathbb{R}^{D}$, the mean squared error (MSE) loss is calculated against the positive extreme $E_{P} \in \mathbb{R}^{D}$ whereas for a negative class MVTS instance $X_{N}^{\text{(k)}} \in \mathbb{R}^{D}$, the MSE loss is calculated against the negative extreme $E_{N} \in \mathbb{R}^{D}$. This approach is designed to facilitate the contrastive learning of MVTS data points such that similar points that belong to the same class will be projected closer around their corresponding extreme points in the new representation space while data points that belong to different classes will be projected further apart. This personalized approach ensures that our embeddings are finely tuned to capture the distinctions between different class instances, facilitating accurate classification. In the end, $|P|$ and $|N|$ being the number of positive and negative class instances respectively, the final objective is to minimize this combined loss of:

\begin{equation}
\begin{aligned}
\mathcal{L}_{\text{Extreme}} = &\frac{1}{|P|} \sum_{k=1}^{|P|} \sum_{i=1}^{D} (X_{P}^{(k)}[i] - E_{P}[i])^2 \\
&+ \frac{1}{|N|} \sum_{k=1}^{|N|} \sum_{i=1}^{D} (X_{N}^{(k)}[i] - E_{N}[i])^2 
\end{aligned}
\label{eq:centroid_reconstruction_loss}
\end{equation}

\subsubsection{Binary Classification of Representation Embeddings}

In the second component of our framework, we use extracted representation embedding vectors $X_{label}^{\text{(k)}} \in \mathbb{R}^{D}$ as input instances to train a downstream classifier to provide a final binary class prediction.   

\section{Experimental Evaluation} \label{sec:Experimental Evaluation}
We present our experimental findings here. The source code for our model and experiments is on our GitHub repository. \footnote{\url{https://github.com/OnurVural/contrex}}

\subsection{Benchmark Dataset} \label{sec:Dataset}

\begin{figure}
\centering
\includegraphics[width = 0.95\linewidth]{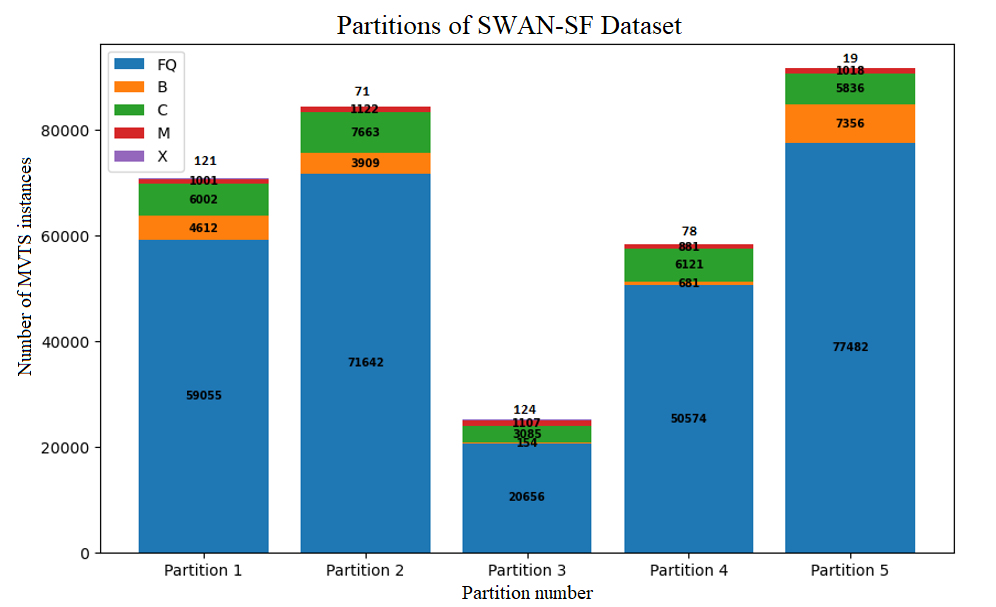}
\caption{ Flare class distribution of each division in stacked bar plot format. For each partition, the count of five flare classes is shown in the plot.}
\label{fig:mvts_partitions}
\end{figure}

SWAN-SF was introduced as an MVTS dataset to facilitate unbiased flare forecasting, and drive advancements in solar flare prediction \cite{angryk2020multivariate, eskandarinasab_2024_11564789, eskandarinasab_2024_11566472}. The data points in SWAN-SF are categorized into five distinct classes representing varying flare intensities. In binary solar flare prediction, we consider positive (P) examples as major flares responsible for health risks and infrastructural damages (i.e., M and X classes) and negative (N) examples as minor flares (i.e., B and C classes) and flare quiet events (i.e., FQ class) \cite{hamdi2017time}. In SWAN-SF, a significant class imbalance exists due to the rarity of major flare events, with N examples greatly outnumbering P examples. This imbalance often results in a bias towards the majority class, yielding high true negative and low true positive rates, complicating the objective of accurately detecting solar flares \cite{ahmadzadeh2021train}. Each data instance in the SWAN-SF dataset is an MVTS slice $M^{\text{(k)}} \in \mathbb{R}^{\tau \times N}$ representing a collection of univariate time series of 1-hour length $\tau$, each having 24 magnetic field parameters \cite{bobra2015solar} as $N$, extracted using a sliding window approach. Each time slice is labeled according to the most intense flare that occurs within the prediction window.  The data is divided into five partitions (i.e., p1, p2, ..., p5), each covering a different observation period \cite{angryk2020multivariate}. The distribution of classes is illustrated in detail in Fig. \ref{fig:mvts_partitions}, underscoring the challenge posed to the primary objective of accurately detecting solar flares in flare-forecasting research.   

\subsection{Performance Evaluation Metrics}  \label{sec:eval}
Due to the substantial P and N class imbalance in SWAN-SF, accuracy alone, which reports the number of correct predictions without going into class specifics, is inadequate. Therefore, we use several additional metrics that are commonly used in previous works of solar flare research for assessing binary solar flare classification: F1 score, receiver operating characteristic area under the curve (ROC AUC) that measures the classifier's ability to distinguish between classes, Heidke Skill Score 2 (HSS2) that evaluates improvement over random predictions, Gilbert Skill Score (GS) that assesses the likelihood of obtaining true positives by chance, and True Skill Statistic (TSS) \cite{hamdi2017time, bobra2015solar, mason2010testing, barnes2008evaluating}. Among these, TSS is particularly robust against class imbalance, expressing the difference between true positive and false positive rates ranging from -1 (all incorrect predictions) to 1 (all correct predictions), with 0 indicating random predictions. TSS is recommended as the primary measure for evaluating solar flare prediction models \cite{bobra2015solar}. The evaluation of classification results, involving true positives (TP), true negatives (TN), false positives (FP), and false negatives (FN), can be expressed as:

\begin{equation}
\text{Accuracy} = \frac{TP + TN}{TP + TN + FP + FN} \label{eq:a}
\end{equation}

\begin{equation}
F1 = \frac{TP}{TP + \frac{1}{2}(FP + FN)} \label{eq:f1}
\end{equation}


\begin{equation}
\resizebox{.9\hsize}{!}{$\text{HSS}_2 = \frac{2 \times [(TP \times TN) - (FN \times FP)]}{[(TP + FN) \times (FN + TN)] + [(TN + FP) \times (TP + FP)]}$} \label{eq:hss2}
\end{equation}

\begin{equation}
\text{GS} = \frac{TP - C}{TP + FP + FN - C} \label{eq:gs}
\end{equation}

\begin{equation}
\text{where } C = \frac{(TP + FP) \times (TP + FN)}{P + N} \label{eq:c}
\end{equation}

\begin{equation}
\text{TSS} = \frac{TP}{TP + FN} - \frac{FP}{TN + FP} \label{eq:tss}
\end{equation}

\subsection{Preprocessing and Training Settings}
For preprocessing the MVTS instances of SWAN-SF, we performed KNN imputation \cite{pujianto2019k} to fill the missing values followed by instancewise normalization of data points across individual time series features. For training and testing purposes, consequent SWAN-SF partitions are used respectively (e.g., p1 for obtaining the extremes and training our framework, p2 for testing). Our framework is trained with the following hyperparameter settings: LSTM with input dimension 24 ($N$ as \# time series in MVTS) and hidden state dimension 128, dropout probability: 0.5, MLP with input size 128 and output size 528 (D-dimension), \# training epochs: 20, Adam learning rate: $10^{-2}$ from the search space of [$10^{-1}$, $10^{-2}$, $10^{-3}$, $10^{-4}$]. 

\subsection{Study of Components in CONTREX}
\begin{figure}
\centering
\includegraphics[width=1.02\linewidth]{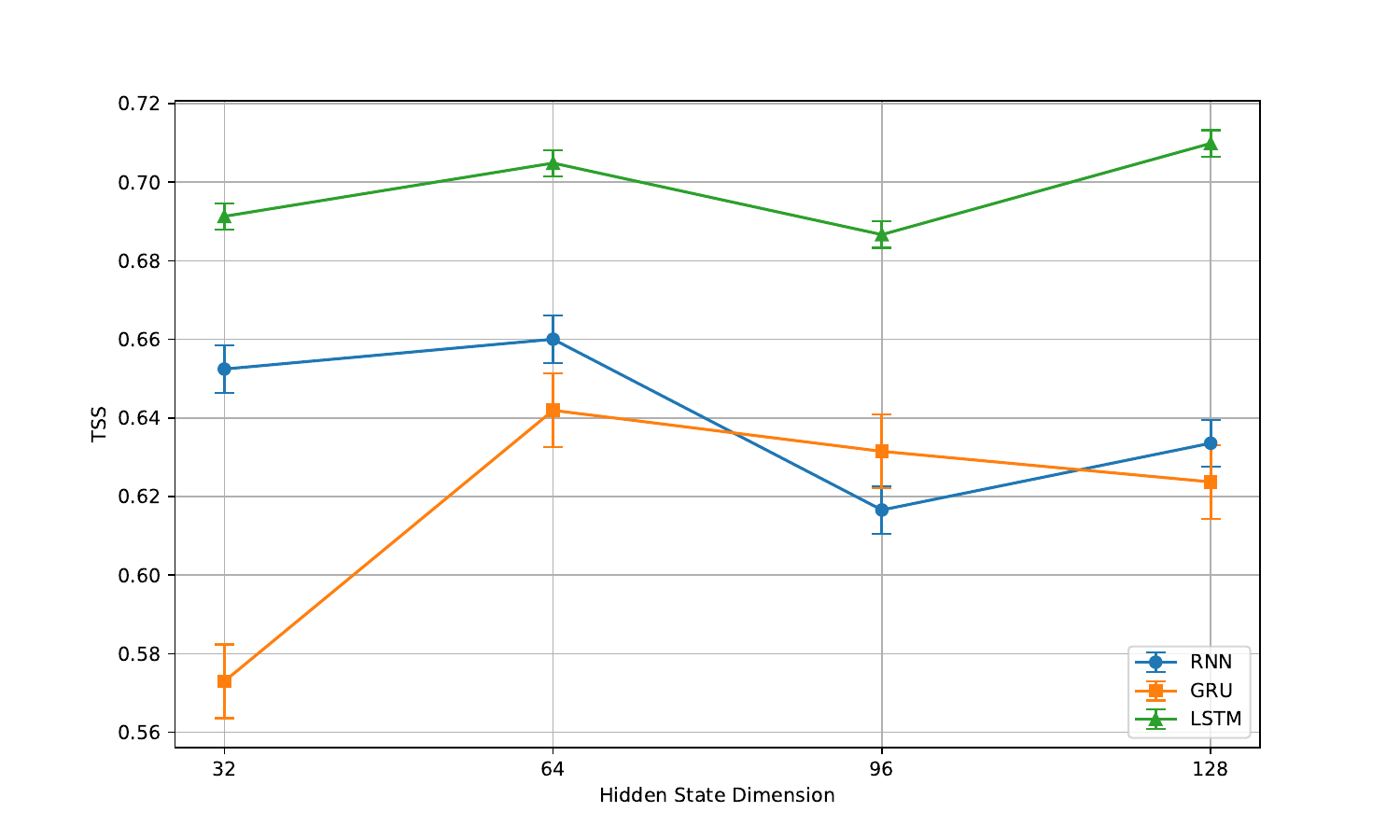}
\caption{Averaged TSS results of RNN, GRU, and LSTM within the CONTREX framework for binary solar flare prediction with varying hidden dimensions.}
\label{fig:roc_hidden}
\end{figure}

We conducted several experiments to evaluate the effects of different components within the CONTREX framework. First, LSTM-based sequence modeling of MVTS data points was compared with other deep learning sequence models, namely recurrent neural network (RNN) and gated recurrent unit (GRU). This experiment also varied the hidden space dimensions to assess their impact on model underfitting or overfitting. Fig. \ref{fig:roc_hidden} shows that the preferred model LSTM exhibits the best average performance in three consecutive partitions as train-test pairs (i.e., p1-p2, p2-p3, and p3-p4) under varying dimensionality in hidden space. Following that, the downstream classifier is selected as logistic regression showing the best average performance in the same three consecutive partitions as train-test pairs as reflected in Table \ref{table:downstream} after experimenting with support vector machine (SVM), k-neighbors classifier (KNC), decision tree (DT), multilayer perception (MLP), and fully-connected network (FC).

\begin{table*}
\centering
\caption{Binary Solar Flare Prediction Performance Results of CONTREX with Various Downstream Classifiers}
\label{table:downstream}
\small
\begin{tabular}{lcccccccc}
\toprule
Classifier & Accuracy & TSS & HSS2 & F1 & GS & ROC AUC \\
\midrule
LR & \textbf{0.7306 ±0.09662} & \textbf{0.7098 ±0.09779} & \textbf{0.1189 ±0.04965} & \textbf{0.1579 ±0.04766} & 0.02303 ±0.007004 & \textbf{0.8549 ±0.04889}\\
SVM & 0.6819 ±0.1344 & 0.6645 ±0.1362 & 0.1026 ±0.055  & 0.1428 ±0.051 & 0.02318 ±0.007065
& 0.8322 ±0.06812\\
KNC	& 0.7163 ±0.125 & 0.693 ±0.1248 & 0.1182 ±0.06089 & 0.1574 ±0.0564 & 0.02297 ±0.007
& 0.8465 ±0.06242\\
DT	& 0.7216 ±0.1354 & 0.6318 ±0.1176 & 0.1113 ±0.05374 & 0.1507 ±0.04774 & 0.02068 ±0.006054 & 0.8159 ±0.05879\\
MLP & 0.6805 ± 0.08599 & 0.6671 ± 0.09098 & 0.09438 ±0.03769 & 0.1348 ±0.03946 & \textbf{0.02322 ±0.006865} & 0.8336 ±0.04549\\
FC & 0.6824 ±0.09883 & 0.6662 ±0.1018 & 0.09238 ±0.02994 & 0.1331 ±0.02902 & 0.0232 ±0.007036 & 0.8331 ±0.0509\\
\bottomrule
\end{tabular}
\end{table*}

\subsection{Representation Embedding Analysis}
\begin{figure}
\centering
\includegraphics[width=0.95\linewidth]{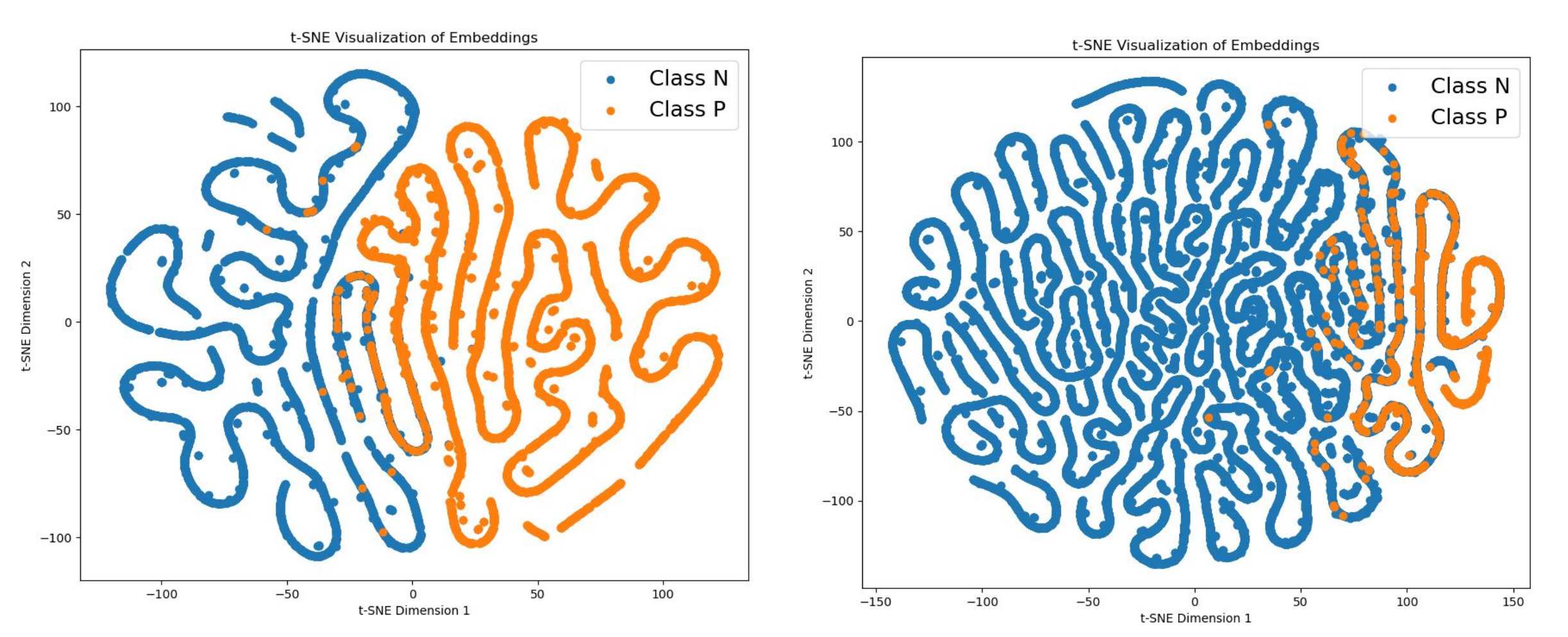}
\caption{t-SNE visualization of partition 3 and partition 4 train-test pair embeddings.}
\label{fig:tsne_centroid}
\end{figure}

To assess the extent of the contrastive abilities of our embeddings in separating classes, t-SNE visualization \cite{van2008visualizing} is performed to map the extracted embeddings $X_{label}^{\text{(k)}} \in \mathbb{R}^{D}$ into two-dimensional space. Fig. \ref{fig:tsne_centroid} gives insight into benchmark dataset's data distribution and suggests that our contrastive learning model effectively separates P and N classes.  

\subsection{Baselines}
To evaluate the performance of our proposed framework, we use the following baselines from current solar flare research where each method obtains a different data representation out of MVTS instances. When a downstream classifier is needed, we use logistic regression to be consistent with our experiments.
\begin{itemize}[leftmargin=*]
\item \textbf{Vector MVTS (VMVTS)}: Flattens each MVTS instance to train a downstream classifier \cite{hamdi2017time}.
\item \textbf{Vector of last timestamp (LTV)}: Extracts magnetic field parameters from only the last timestamp of the MVTS being temporally nearest to the flaring event to train a downstream classifier \cite{bobra2015solar}.
\item \textbf{Long-short term memory (LSTM):} MVTS instances are fed into LSTM as a timestamp vector at each time step, and the last hidden representation is extracted as the representation \cite{muzaheed2021sequence}. We maintain the same hyperparameter settings: a hidden state dimension of 128, and a stochastic learning rate of $10^{-2}$. 
\item \textbf{Random convolutional kernel transform (ROCKET):} Leverages the strengths of convolutional neural networks for time series classification by using randomly generated convolutional kernels, achieving state-of-the-art accuracy on 26 MVTS datasets of the UEA archive \cite{dempster2020rocket, ruiz2021great}.
\end{itemize}

\subsection{Binary Classification Performance}

\begin{figure}
\centering
\includegraphics[width=0.94\linewidth]{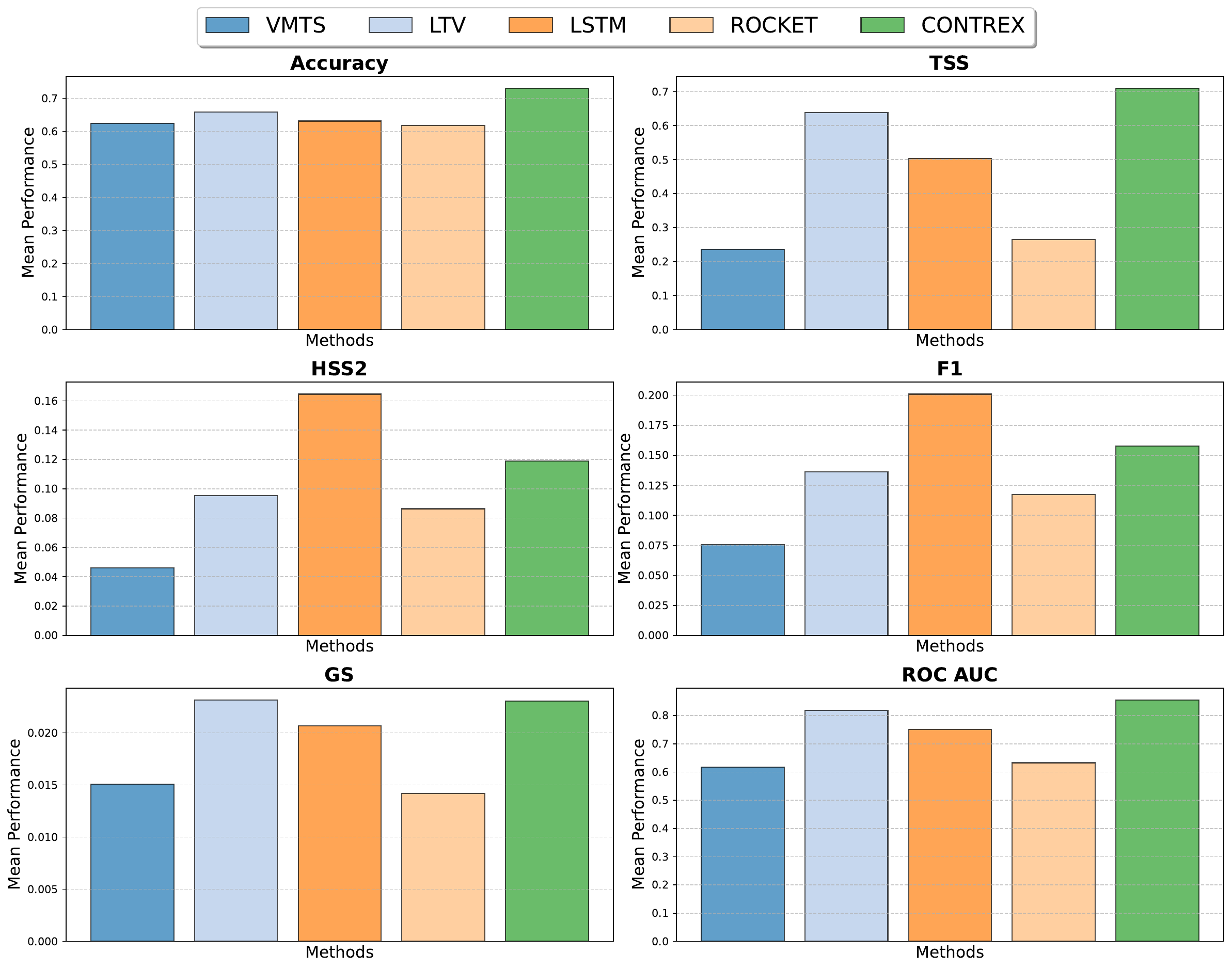}
\caption{Binary solar flare classification performance of baselines. }
\label{fig:performance}
\end{figure}

Following our experiments with three consecutive partitions as train-test pairs (i.e., p1-p2, p2-p3, and p3-p4), the averaged performance results are displayed in Table \ref{table:comparison} and Fig. \ref{fig:performance}, demonstrating that the contrastive ability of CONTREX manages to show promising performance. CONTREX emerges as the top performer with 0.7306 accuracy, 0.7098 TSS and 0.8549 ROC AUC, outperforming LTV by 5.1\%, 7.2\% and 3.6\%, respectively. However, with 0.1189 HSS2 and 0.1579 F1 score, CONTREX has the second position, lagging behind the leading performer, LSTM, by 4.6\% and 4.3\%, respectively. With 0.02303 GS, CONTREX shares the top position with LTV, with only a marginal 0.011\% difference between the two. Overall, our experimental findings underscore CONTREX's competitiveness against state-of-the-art methods.

\begin{table*}
\centering
\caption{Binary Solar Flare Prediction Performance Results of Data Representation Methods}
\label{table:comparison}
\small
\begin{tabular}{lcccccc}
\toprule
Model & Accuracy & TSS & HSS2 & F1 & GS & ROC AUC \\
\midrule
VMTS & 0.6243 ±0.4204 &	0.2358 ±0.3335 & 0.04592 ±0.06494 & 0.07575 ±0.06675	& 0.01507 ±0.01333 & 0.6179 ±0.1667\\
LTV	& 0.6589 ±0.1728 & 0.6378 ±0.1748 & 0.09523 ±0.05173 & 0.1361 ±0.04493 
& \textbf{0.02314 ±0.007139} & 0.8189 ±0.08738\\
LSTM & 0.6314 ±0.2914 & 0.5027 ±0.1967 & \textbf{0.1646 ±0.1766} & \textbf{0.2009 ±0.1626} & 0.02068 ±0.008318 & 0.7514 ±0.09834 \\
ROCKET & 0.6174 ±0.2688 & 0.2652 ±0.162 & 0.08639 ±0.07387 & 0.1174 ±0.06113 & 0.01416 ±0.01185 & 0.6326 ±0.08102  \\
CONTREX & \textbf{0.7306 ±0.09662} & \textbf{0.7098 ±0.09779} & 0.1189 ±0.04965 & 0.1579 ±0.04766 & 0.02303 ±0.007004 & \textbf{0.8549 ±0.04889}\\
\bottomrule
\end{tabular}
\end{table*}

\section{Conclusion} \label{sec:Conclusion}
In this paper, a novel contrastive approach for time series is introduced and evaluated by the task of binary solar flare prediction of the SWAN-SF data instances. Our methodology comprised extracting dynamic attributes from each MVTS instance, computing contrastive extreme points from feature vectors, obtaining sequence representation embeddings for MVTS data instances guided by our custom reconstruction loss that leveraged the idea of generating embeddings that encapsulate the distinctive class characteristics, and training a downstream classifier with embeddings to binary classify solar flares. In future studies, we aim to enhance the performance of our framework. For this reason, bringing other elements of triplet loss to our loss function as commonly utilized in modern contrastive learning methods, experimenting with different sampling and extreme point calculation methodologies, and applying feature reduction to the extreme points will be possible study directions. Furthermore, a study is planned to test the abilities of our framework in its current state with different MVTS benchmark datasets having binary and multi-class conditions. This experimentation will help to understand whether utilizing the contrastive extremes can yield better separation results for other time series settings.   

\section*{Acknowledgment}
This project has been supported in part by funding from CISE and GEO Directorates under NSF awards \#2204363, \#2240022, \#2301397, and \#2305781.

\bibliographystyle{IEEEtran}
\bibliography{conference_contrex_icmla.bib}

\end{document}